\documentclass[11pt]{article}
\usepackage{epsfig}
\newcommand{\sss}{\vspace{.2in}}
\newcommand{\be}{\begin{equation}}
\newcommand{\ee}{\end{equation}}
\newcommand{\sn}{{\rm sn}}
\newcommand{\dn}{{\rm dn}}
\newcommand{\cn}{{\rm cn}}
\newcommand{\sech}{{\rm sech}}

\begin{document}
\sss\sss\begin{center}
{\Large \bf   
Relating Linearly Superposed Periodic Solutions of Nonlinear Equations to One Soliton Solutions}
\end{center}
\vspace{0.5 in}
\begin{center}
{\large{\bf
   \mbox{W. Reinhardt}$^{a}$,
   \mbox{A. Khare}$^{b}$,
   \mbox{U. Sukhatme}$^{c}$
 }}
\end{center}
\vspace{0.5 in}
\noindent
$^{a}$
Departments of Chemistry and Physics, University of Washington,
Seattle, WA 98195-1700\\
$^{b}$
Institute of Physics, Sachivalaya Marg, Bhubaneswar 751005, India\\
$^{c}$
Department of Physics, State University of New York at Buffalo,  
Buffalo, NY 14260 \\

\vspace{1.5 in}  
{\bf {Abstract.}}
Using new generalized Landen transformations, we prove that the solutions
of the KdV and other nonlinear equations obtained recently by using a kind of
superposition principle for periodic solutions are in fact novel re-expressions of
well known periodic one soliton solutions. \\

\newpage

Nonlinear partial differential equations have a variety of unusual and 
sometimes unexpected solutions. For instance, it was not realized \cite{1} 
until 1965 that the Korteweg-de Vries (KdV) 
equation \cite{2}
\be\label{1}	
			u_t  - 6uu_x + u_{xxx} = 0~,	
\ee
supported multi-solitonic solutions, in addition to the single solitary wave 
solutions it was developed in 1895 to describe \cite{3}.   
However, for traveling wave solutions of the form $u(x,t) = u(x-ct)$, the KdV 
equation has an 
integrating factor, and reduces to a nonlinear ordinary differential 
equation, immediately integrable to yield a Weierstrass function solution. 
In terms of the standard Jacobian elliptic functions 
$\sn(x,m),\,\cn(x,m),\,\dn(x,m)$ \cite{4} 
with argument $x$ and modulus parameter $m$, a specific range of integration 
constants corresponds to the well-known cnoidal type periodic solution
\be\label{2}
u_1 (x,t) = -2 \alpha^2 \dn^2 [\alpha (x- b_1 \alpha^2 t),{m}] + \beta \alpha^2 ~,
\ee
with velocity $b_1 = 8-4{m}-6\beta$.
It is worth noting that in the limit $m \rightarrow 1$, the period becomes 
infinite, and
solution (2) reduces to the single soliton solution
$-2\alpha^2 \sech^2 [\alpha (x-b_1\alpha^2 t)] + \beta \alpha^2$.  
  
In a recent paper, Khare and Sukhatme \cite{5} pointed out that certain kinds 
of linear superposition of the periodic solutions given in Eq. (\ref{2}) are 
also solutions of the KdV equation by virtue of several new cyclic 
identities \cite{6} satisfied by the Jacobi elliptic functions. 
The solutions 
are of the form
\be\label{3}
u_p(x,t) = -2\alpha^2 \sum_{i=1}^{p}  d_i^2  + \beta \alpha^2~,			
\ee
where $p$ is any integer and
\be\label{4}
d_i \equiv \dn[\alpha (x-b_p \alpha^2 t) +\frac{2(i-1)K(m)}{p},m]~.
\ee
The velocity is $b_p = 8-4{m}-6\beta+12A(p,m)$ with $A(p,m)$ being a 
constant in a certain cyclic identity \cite{5}.

In addition to the KdV equation, linear combination solutions similar to 
Eq. (\ref{3}) were also shown to exist \cite{7} for many other nonlinear 
differential equations. In the context of the static sine-Gordon equation, it 
was subsequently pointed out \cite{8} that such linearly superposed solutions 
cannot be new, but must be re-expressible in terms of the initial periodic 
solution with different modulus parameter and integration constants. This 
connection was established by the discovery of new generalized Landen 
transformations \cite{8,9}.  	
The point of this letter is to demonstrate that for the KdV equation also, 
the periodic solutions of the form (\ref{3}), although they appear to be new, can be simply 
re-expressed in terms 
of the original single Jacobi function solution given by Eq. (\ref{2}).
 
Our procedure consists of starting with the recently found generalized Landen 
formula for $\dn(x,\tilde{m})$ which involves a transformation from a modulus 
parameter $m$ to a new modulus parameter $\tilde{m}$ \cite{8,9}: 
\be\label{5}
\dn(x,\tilde{m}) = \gamma \left[\dn(\gamma x,m)+\dn(\gamma x +2K(m)/p,m)
+...+\dn(\gamma x+2(p-1)K(m)/p,m)\right]~,
\ee
where
\begin{eqnarray} \label{6}
\gamma &=& [1+\dn(2K(m)/p,m)+...+\dn(2(p-1)K(m)/p,m)]^{-1}~,\nonumber\\
{\tilde m} &=& (m-2) \gamma^2+2\gamma^3 [1+\dn^3(2K(m)/p,m)+...+\dn^3(2(p-1)K(m)/p,m)]~.
\end{eqnarray}
Squaring both sides yields a Landen transformation for $\dn^2(x,\tilde{m})$:
\be \label{7}
\dn^2(x,\tilde{m})= \gamma^2\left[ \sum_{i=1}^{p} \dn^2\, [\gamma x +2(i-1)K(m)/p,m]+\sum_{r=1}^{p} a_p(r)  \right]~,
\ee
where the constants $a_p(r)$ come from the cyclic identity 
$a_p(r)\equiv(d_1d_{1+r}+{\rm cyclic~permutations})$ \cite{6}.
It now follows that solution (\ref{3}) can be written as 
$-2 {\tilde \alpha}^2 \dn^2 [{\tilde \alpha}(x- {\tilde c}t),{\tilde m}]  
+ {\tilde \beta} {\tilde \alpha}^2$, provided one makes the identification
\be\label{8}
{\tilde \alpha} = \alpha/\gamma~,~~{\tilde c} = b_p \alpha^2~,~~
{\tilde \beta} = \beta \gamma^2+2\gamma^2\sum_{r=1}^{p} a_p(r)~.
\ee

Two other periodic traveling wave solutions of the KdV equation are
\be\label{9}
{u}_{\pm} (x,t) = 
\alpha^2 \left[{m}\sn^2(\eta,{m})
\pm \sqrt{{m}}\cn(\eta,{m})\,\dn(\eta,m)\right]~,
~~\eta \equiv \alpha(x-q_1 \alpha t)~,
\ee
with velocity $q_1= -1-{m}$. These solutions are different from Eq. (\ref{2}) 
since they correspond to choosing a different range of integration constants 
in the general Weierstrass function solution of the KdV equation. They can 
also be obtained by making a Miura transform of the modified KdV equation.  
Linear superposition again gives apparently new solutions of the KdV equation, 
but in fact that is not the case. The solutions can be re-written in terms of 
the original solution (\ref{9}) albeit with a change of parameters, using the 
generalized Landen 
transformation for $\sn(x,\tilde{m})$ and $\cn(x,\tilde{m})$ \cite{8}, with 
the same $\tilde{m}, m$ connection given in Eq. (\ref{6}). A similar comment 
also applies to other nonlinear differential equations. Thus, although the idea of 
judicious linear superposition of periodic solutions has not given new 
solutions for nonlinear equations, it has certainly yielded 
many novel, previously unexpected identities among  the Jacobian elliptic 
functions themselves, some involving cyclic combinations \cite{6} and others containing 
modulus parameter transformations \cite{8,9}.

W.P.R. acknowledges the partial support of the US National Science Foundation,
Grant PHY-0140091; and from NIST-NSF Digital Library of Mathematical Functions 
Project; U.S. thanks the U.S.  
 Department of Energy for partial support of this research 
under grant DOE FG02-84ER40173. \\

\end{document}